\documentstyle[prd,aps,preprint,graphics]{revtex}
\begin{document} 
\newcommand{\cL}{{\mathcal{L}}}  
\title{Constraints on Variant Axion Models} 
\author{Mark Hindmarsh\cite{mhaddress}}
\author{Photis Moulatsiotis\cite{pmaddress}} 
  \address{Centre for Theoretical Physics \\
  University of Sussex \\ Brighton BN1 9QJ \\ U.K} 
\date{August 1997}
\preprint{SUSX-TH-97-014, hep-ph/9708281}
      
\maketitle
\begin{abstract}
\setlength{\baselineskip}{14pt}
 A particular  class of variant  axion  models with two higgs  doublets and a
  singlet  is  studied.  In these  models  the  axion  couples  either  to the
  $u$-quark  or  $t$-quark or both, but not to $b$, $c$, $s$, or $d$. When the
  axion couples to only one quark the models possess the desirable feature of
  having no domain wall problem, which makes them viable candidates for 
  a cosmological axion string scenario. We 
  calculate the axion  couplings  to  leptons,
  photons  and  nucleons, and  the  astrophysical
  constraints on the axion decay constant $v_a$ are investigated and 
  compared to the DFSZ axion model.
  We find that the most restrictive lower bound on $v_a$, that from SN1987a,
  is lowered by up to a factor of about 35, depending on the model and also the 
  ratio of the vacuum expectation values of the higgs doublets. 
  For scenarios with axionic strings, the allowed window for $v_a$ 
  in the $u$ quark model can be more than two orders of magnitude. 
  For inflationary scenarios, the cosmological
  upper bound on $v_a/N$, where $N$ is the QCD anomaly factor, 
  is unaffected: however, the variant models have $N$ either 3 or 6 times
  smaller than the DFSZ model.
\end{abstract}
\section{Introduction}
  
  The relevance of instantons \cite{1} to physics was first shown by 't Hooft
  \cite{2} in his resolution of the $U_A(1)$  problem  \cite{3}.  It was first
  pointed  out by him that the  topological  structure  of the  vacuum  of any
  non-abelian  gauge theory, which includes QCD, is  non-trivial.  
  There are gauge  transformations, 
  characterized  by  different   topological  numbers  $n$ (the  
  Pontryagin  index), which can not be continuously  deformed to one another.
  This gives rise to distinct  ground  states,  labelled by different  $n$ and
  separated by finite energy barriers.  
  Instantons  can be
  physically  interpreted as quantum mechanical tunneling events, in Euclidean
  spacetime,  between these different  ground  states \cite{4,5}.  
  One then has to construct the true, 
  gauge-invariant,  vacuum out of these degenerate  $n$-vacua by taking 
  a linear combination 
  \begin{equation}
  |\theta\rangle=\sum_n e^{-in\theta}|n\rangle.
  \end{equation} 
In the path integral approach, the effect is to add the so-called 
$\theta$-term, to the ordinary QCD lagrangian, or
  \begin{equation}
  \cL_\theta=\theta{g_{s}^2\over 32\pi^2}G_{b}^{\mu\nu}{\tilde{G}_{b\mu\nu}}.
  \end{equation}
  Under the combined  action of charge  conjugation and parity  transformation
  the  $\theta$-term  changes sign and hence it violates CP  invariance. 
 There is also CP violation communicated from the quark mass matrix $M$: 
 if we diagonalise it with a bi-unitary transformation, we find that 
 the coupling constant $\theta$ is modified to
  \begin{equation}
\bar\theta=\theta+\arg(\det  M). 
  \end{equation}
Despite being the coupling constant of a total derivative term, the 
parameter $\bar\theta$ is observable, through its
  effect  on  the  neutron  electric  dipole  moment   \cite{6}. The  
current experimental upper limit \cite{7} on the electric  dipole  moment  
implies that $\bar\theta$  is less than about $10^{-9}$.  
  
Such a small  value of  $\bar\theta$
  contradicts our expectation that a dimensionless free parameter should be of
  order one.  Many ideas have emerged in trying to  resolve  the  puzzle.
  One of the most elegant  solutions  was proposed by Peccei and Quinn in 1977
  \cite{8}.  The authors postulated the invariance of the lagrangian under the
  transformations  of a new extra global chiral $U(1)$ symmetry, 
  called  PQ-symmetry,
  thus   enlarging   the   symmetry   group   of  the  SM  to   $SU(3)_C\times
  SU(2)_{L}\times U(1)_Y\times U(1)_{PQ}$.  To accomodate the extra charges of
  the new  symmetry  one  needs (at least) one  extra  higgs  doublet.  
  When  this   symmetry  is
  spontaneously broken, a pseudoscalar boson appears \cite{9} in the
  theory,  called the {\it  axion}.  Normally,  one  would think  
  that the axion is   massless, as it is the Nambu-Goldstone boson of
  the PQ symetry.  However, the  $U(1)_{PQ}$  symmetry is an
  anomalous one, spoiled by the effect of instantons in the QCD vacuum, and
  the axion picks up a small mass via the axion-gluon-gluon triangle
  anomaly.

  The   assignment  of   appropriate   PQ-charges  to  the  higgs  fields  and
  consequently   to  the  quarks  is  responsible  for  the  presence  of  the
  anomaly in the PQ current,  and also
  for the variety of the axion  models.  In the original  model,  known as the
  PQWW (Peccei-Quinn-Weinberg-Wilczek) axion model  \cite{9},  
  all the quarks of the same chirality were  assigned the same PQ charge.  
  Unfortunately  the model  was  ruled out both by  particle   physics
  experiments and  by  astrophysical observations.  The former gave constraints
  which came from $K,\ J/\psi$ and $\Upsilon$ meson decays, reactor and
  beam dump  experiments  and nuclear  deexcitations  \cite{10}. The latter
  are   in    fact    more    restrictive    and    imply    that
  $v_a>10^{10}$ GeV for most axion models considered to date 
  \cite{11}.  A way out of this was first proposed by
  Kim in 1979 and  subsequently by Shifman,  Vainstein \& Zakharov  \cite{12},
  who added a higgs singlet, thus enabling the axion decay constant to be much
  higher than the  electroweak  scale.  However, it cannot be too high, as 
  it is also possible to restrict  $v_a$
  from  above through  cosmological arguments.  
  Coherent  oscillations  in axion  field can be
  produced  after  inflation  or via the  formation of 
  axionic cosmic  strings.  
  The   requirement   that  the  energy density in these oscillations 
  is not large enough to overclose the  universe puts an 
  upper limit on $v_a$.  The current
  values on these  limits for the axion decay  constant  are  $v_a$ 
  less than about $10^{12}$GeV
  \cite{13,11} from inflationary scenarios and $v_a\lesssim 2.6\times 
  10^{11}$GeV  \cite{14} from axion  strings (for $H_0 = 50$ 
  km s$^{-1}$ Mpc$^{-1}$, and making the conservative assumption 
  that radiation from infinite strings
  dominates that from loops).  As we can see there is only a very small  window 
left  for the axion.
  
  Cosmology also restricts on the value of $N$, the parameter that characterises
  the QCD anomaly, which is related to the number of quarks that couple to the 
  axion.  If there is no inflation 
  between the PQ symmetry-breaking transition and the present day, a 
  dense network of axion strings is formed. At around a temperature of
  1 GeV, each string becomes the junction of (in our normalisation convention) 
  $2N$ domain walls
  \cite{15}. In  order to avoid the domain walls dominating the energy density,
  $2N$ must be equal to unity, so that the string-wall system can annihilate.
  Models with $N>{1\over 2}$ must have a period of inflation at a low energy 
  scale, or must reheat after inflation to less than the PQ symmetry-breaking 
  temperature, to remain viable.

In this paper we examine some variant axion models based on those proposed  
  first by Peccei, Wu \& Yanagida  \cite{16}  and  independently  by Krauss \&
  Wilczek  \cite{17}.  Their models were constructed  
  with two higgs doublets and assigned
  of different PQ charges to different quarks.  The original 
  reason for this is that in
  order to make an axion model which avoided the particle physics constraints 
  at the time it was essential  to decrease  the axion
  couplings  to $c$ {\it  and} $b$  quarks  on one  hand,  and on the  other to
  sufficiently  suppress the  $K^{+}\rightarrow  \pi^+a$ decay rate.  One must
  also have a  sufficiently  short lived axion so it cannot be detected by the
  other  experiments.  To accomplish  that, one has to couple both the $c$ and
  $b$ quarks to the same higgs  field.  However,  as we know the limits on the
  strangeness  changing neutral currents are very tight and for that reason we
  have to couple the strange  quark to the same higgs  doublet as $c$ and $b$.
  Thus the only way to realise the  Peccei-Quinn  symmetry is to couple either
  the $u$ or the $t$ or both  quarks  to a second  higgs  doublet.  So we have
  three different models in our hands, two of which have the 
  cosmologically desirable property that  $N={1\over 2}$, and hence have no 
  domain wall problem.
  
  Although the original  models are also ruled out, along with the PQWW axion,
  by the  astrophysical  constraints,  they also have extensions  with a higgs
  singlet.  The  astrophysical  constraints on these models,  presented in the
  next  section,  have not  previously  been  considered,  and we  present  an
  analysis in this paper of the bounds on the axion scale that can be inferred
  from their  couplings to electrons,  photons, and nucleons in  astrophysical
  processes.  We find, as for the standard DFSZ  \cite{18} and KSVZ  \cite{12}
  invisible  axions,  that the  tightest  constraint  comes,  via the  nucleon
  coupling,  from SN1987a.  However, in these variant  models, the coupling is
  generally  weaker,  and  weakens  the  lower  bound on $v_a$ by a factor  of
  between 1.4 and 35, depending on which quarks the axion  couples to, and on 
  the ratio  of the  vacuum  expectation  values  of the  higgs  doublets. The
  cosmological upper bound on $v_a$ from the axion density is also reduced, by
  a factor of either 3 or 6, as a result of the smaller QCD anomaly factor. 

\section{Description of the model}

  The models we are going to discuss were first proposed by Geng \& Ng \cite{19}  
  and has elements  from both the  previously  discussed  one and that of DFSZ.
  There, an extra higgs singlet $\phi$ is introduced, the phase
  of which is reserved  for the axion.  A direct  coupling of $\phi$ to quarks
  and leptons is impossible; but in the DFSZ model it couples to both $\phi_1$
  and  $\phi_2$.  We however  couple  $\phi$ only to one of the two  doublets,
  namely $\phi_1$, which then couples to the `special' quarks according to the
  three models  discussed above.  The other higgs field couples to the rest of
  the  quarks  and  the  leptons.  Therefore  we  have  the  following  Yukawa
  interactions:
  \newcounter{c}\setcounter{c}{1}
  \renewcommand{\theequation}{{\textrm{{Model} \Roman{c}}}}
  \begin{eqnarray}
  \cL_Y&=&f_{i1}^u(\bar q_{Li}\phi_1 u_{R1})+f_{ij}^d(\bar q_{Li}\phi_2
  d_{Rj})+\sum_{j=2,3} 
     f_{ij}^u(\bar q_{Li}\tilde{\phi}_2 u_{Rj})+h.c.\\\addtocounter{c}{1}
  \cL_Y&=&f_{i3}^u(\bar q_{Li}\phi_1 u_{R3})+f_{ij}^d(\bar q_{Li}\phi_2
  d_{Rj})+\sum_{j=1,2}
     f_{ij}^u(\bar q_{Li}\tilde{\phi}_2 u_{Rj})+h.c.\\\addtocounter{c}{1}
  \cL_Y&=&f_{i1}^u(\bar q_{Li}\phi_1 u_{R1})+
   f_{i3}^u(\bar q_{Li}\phi_1 u_{R3})+f_{ij}^d(\bar q_{Li}\phi_2
  d_{Rj})+f_{i2}^u(\bar q_{Li}\tilde{\phi}_2 u_{R2})+h.c.
  \end{eqnarray} where
\addtocounter{equation}{-3}
\renewcommand{\theequation}{\arabic{equation}}
$\tilde \phi = i\sigma_2\phi^*$.  Our nomenclature is
  the same as that of Peccei, Wu \&  Yanagida  \cite{16}.  In the
  first  model it is the $u$ quark that couples to
  the axion, whereas in the second is the $t$.  Finally in the last model both
  quarks couple to the axion.

  The most general renormalizable  potential for the model, consistent with 
  gauge, as well as PQ symmetry and renormalisability is \cite{12,19}
  \begin{eqnarray}  
  V & = & \lambda_1(\phi_1^{\dag}\phi_1-v_1^2/2)^2+\lambda_2(\phi_2^{\dag}
  \phi_2-v_2^2/2)^2+\lambda(\phi^{\ast}\phi-v^2/2)^2+\nonumber \\ 
    &   & (a\phi_1^{\dag}\phi_1+b\phi_2^{\dag}\phi_2)\phi^{\ast}\phi+c(\phi_1^T
  i\sigma_2\phi_2\phi^2+h.c)+d|\phi_1^T i\sigma_2\phi_2|^2+
  e|\phi_1^{\dag}\phi_2|^2
  \end{eqnarray}
  The appropriate PQ transformations according to which the quarks acquire a PQ
  charge, as well as leaving the Yukawa lagrangians invariant are
  \begin{eqnarray}
  q_{Rj} & \longrightarrow & e^{iQ_{Rj}\alpha}q_{Rj} \nonumber \\
  q_{Lj} & \longrightarrow & e^{iQ_{Lj}\alpha}q_{Lj} \\
  \phi_n & \longrightarrow & e^{i(Q_{Lj}-Q_{Rj})\alpha}\phi_n \nonumber 
  \end{eqnarray}
  where $\alpha$ is explained below. 
  In the models under discussion, we can choose an assignment of 
  PQ charges such that $Q_{Lj}=0$ and only some of the $Q_{Rj}\neq 0$.
  Furthermore, we impose the normalisation condition that $Q_{Rj}=1$
  for the `special' quarks, which are $u_R$ and/or $t_R$. That 
  further fixes the transformations for the higgs fields, which are as follows:
  \begin{eqnarray}
  \langle\phi_1\rangle &=& {1\over\sqrt{2}}v_1\exp{[-i(\alpha+\alpha_Z)]},
  \nonumber \\
  \langle\phi_2\rangle &=& {1\over\sqrt{2}}v_2\exp{(i\alpha_Z)}, \\
  \langle\phi\rangle &=& {1\over\sqrt{2}}v\exp{(i\alpha)}, \nonumber
  \end{eqnarray}
  where $v_1$, $v_2$, $v$ are the vacuum expectation values of the higgs fields
  and $\alpha$, $\alpha_Z$ are the angles conjugate to the axion and the 
  longitudinal degree of freedom of the $Z^0$ boson respectively.
  
  Our following step is to find an expression  for the axion decay  constant. 
  Let $a^\prime$  and $Z$ be the two  goldstone  bosons  after the breaking of 
  the  Peccei-Quinn symmetry and before instanton effects are taken into
  consideration.  The first one, $a^\prime$, is the massless axion and the
  second, $Z$, the goldstone boson that is eventually  eaten by the $Z^0$.  
  One then has the following equation
  \begin{equation}
  \left( \begin{array}{c}a^\prime \\ Z \end{array}\right)=\left(   
  \begin{array}{lr} v_a & 0 \\ v_{10} & v_{11} \end{array}\right)\left(  
  \begin{array}{c}\alpha \\ \alpha_Z \end{array}\right)
  \label{e:mixmat}
  \end{equation}
  The  $2\times  2$  matrix on the right hand side of (\ref{e:mixmat})
 is the most
  general matrix compatible with the requirement not to mix the axion with the
  $Z^0$  boson.  On the other  hand if one takes the  kinetic  term for  these
  massless scalar degrees of freedom then one has
  \begin{eqnarray}
  {1\over 2}[(\partial_{\mu}a^\prime)^2+(\partial_{\mu}Z)^2] &=& {1\over   
  2}\{(v_1^2+v^2)(\partial_{\mu}\alpha)^2+(v_1^2+v_2^2)(\partial_{\mu}\alpha_Z)
  ^2\nonumber \\ & &
  +v_1^2[(\partial_{\mu}\alpha)(\partial^{\mu}\alpha_Z)+(\partial^{\mu}\alpha)
  (\partial_{\mu}\alpha_Z)]\}.
  \label{e:scakin}
  \end{eqnarray}
  The  derivative of (\ref{e:mixmat}) compared with (\ref{e:scakin}) 
  gives the expression for the axion
  decay constant as well as the electroweak breaking scale
  \begin{eqnarray}
  v_a &=& \sqrt{v^2+{{v_1^2v_2^2}\over {v_1^2+v_2^2}}}, \\
  v_{EW} &=& v_{11}=\sqrt{v_1^2+v_2^2}=\textrm {246 GeV}, \\
  v_{10} &=& {v_1^2\over{\sqrt{v_1^2+v_2^2}}}.
  \end{eqnarray}
  The expression is the same for all three models and as we can see, it is $v$
  that  ultimately  fixes the scale for $v_a$. 
  
 On the other hand if we apply the same normalisation convention to the DFSZ
axion, that is if we assign $Q_{Rj}=1$ to every quark, then the transformation 
of the  higgs fields are:
  \begin{eqnarray}
  \langle\phi_1\rangle &=& {1\over\sqrt{2}}v_1\exp{[-i(\alpha+\alpha_Z)]},
  \nonumber \\
  \langle\phi_2\rangle &=& {1\over\sqrt{2}}v_2\exp{[i(-\alpha+\alpha_Z)]}, \\
  \langle\phi\rangle &=& {1\over\sqrt{2}}v\exp{(i\alpha)}, \nonumber
  \end{eqnarray}
where
\begin{eqnarray}
  v_a &=& \sqrt{v^2+{{4v_1^2v_2^2}\over {v_1^2+v_2^2}}}, \\
  v_{EW} &=& v_{11}=\sqrt{v_1^2+v_2^2}=\textrm {246 GeV}, \\
  v_{10} &=& {{v_1^2-v_2^2}\over{\sqrt{v_1^2+v_2^2}}}.
  \end{eqnarray}
  Our choice of charges ensure that $v_a$, which is essentially equal 
  to the PQ symmetry-breaking scale, is also equal to the axion decay
  constant $f_a$ (as defined by Srednicki \cite{21}). We have avoided 
  using $f_a$ because of its many definitions in the literature.

\section{The axion couplings}

 Our next step is to  determine
  the  couplings  of the axion to  different  quarks  and  leptons.  
In the basis where quarks $q'_j$ are eigenstates of the weak interaction, 
the QCD part of the lagrangian is
  \begin{equation}
  \cL_{QCD}={\bar{q}'_j}(i\gamma_{\mu}D^{\mu})q'_j-
  {\bar{q}'_{Li}}M_{ij}q'_{Rj}- {\bar{q}'_{Ri}}M^{\dag}_{ij}q'_{Lj} - {1\over  
4} G_b^{\mu\nu}G_{b\mu\nu}+\theta{g_{s}^2\over
  32\pi^2}  G_b^{\mu\nu}\tilde    {G}_{b\mu\nu},
  \end{equation}
 where $M$ is the quark mass matrix, and depends on $\alpha$ and $\alpha_Z$,
 and hence $a'$.  If we diagonalise the quark mass matrix, we find that in
 the mass eigenstate basis the lagrangian is
  \begin{eqnarray}
  \cL_{QCD}&=&{\bar{q}_j}(i\gamma_{\mu}D^{\mu}-m_j)q_j-{1\over 4}
  G_b^{\mu\nu}G_{b\mu\nu}+{1\over 2}(\partial_{\mu} a')^2+{g_{s}^2\over32\pi^2}
  (\bar\theta-2N\frac{a'}{v_a})G_b^{\mu\nu}\tilde{G}_{b\mu\nu}-\nonumber  \\ 
  & &-{g_j\over 2v_a}\partial^\mu a'\bar{q}_j\gamma_\mu\gamma_5 q_j,
  \label{e:LQCD}        
  \end{eqnarray}
  where $N$ is explained below.
If there were no mixing between the axion and the $Z^0$, the coupling 
$g_j$ would just be the difference of the PQ charges of the right and 
left-handed quarks $q_j$.  The presence of the mixing modifies the
relation to 
  \begin{equation}
  g_j=(Q_{Rj}-Q_{Lj})-(Y_{Rj}-Y_{Lj}){v_{10}\over v_{11}},
  \end{equation}
  where $Y_{Rj}$ and $Y_{Lj}$ are the hypercharges of the right and left chiral
  fields respectively of the $j$-th quark. 

The constant $N$ depends on the number and type of particles with the 
gluon anomaly
\begin{equation}
  N = \sum_j (Q_{Rj}-Q_{Lj})t_j,
\end{equation}
  where $t_j$ is the index of the $SU(3)_C$ representation to which the fields
  belong (for the known quarks $t={1\over  2}$). Hence, in our normalisation 
  convention, for  Models I and II, $N={1\over 2}$, and for Model
   III, $N=1$.  As advertised, the first two models have no axionic
domain wall problem. In the case of the DFSZ axion we have $N=3$.

When calculating the axion-quark interactions in practice, the 
derivative interaction in (\ref{e:LQCD}) can prove troublesome, 
and it is usual to leave the phases corresponding to the axion
degree of freedom in the quark mass matrix, so that the interaction
term is
  \begin{equation}
  \cL_{aq}=-{m_j }\bar q_je^{i\gamma_5(\Delta Q_j\alpha + \Delta Y_j \alpha_z)} 
  q_j,
  \end{equation}
where $\Delta Q_j=Q_{Rj}-Q_{Lj}$ and $\Delta Y_j=Y_{Rj}-Y_{Lj}$.
To  first   order  in   $a'/v_a$, this gives
  \begin{equation}
  \cL_{aq}=-ig_j\frac{m_j }{v_a} a'\bar q_j\gamma_5  q_j.
  \end{equation}
The (pseudoscalar) Yukawa coupling of the interaction between the 
axion and  the $j$-th quark is then clearly 
 \begin{equation}
  h_{aj} = g_j{m_j \over v_a}.
  \end{equation}
The interactions with leptons can be calculated similarly. The term
involving the axion and the $l$-th lepton $e_l$ is
  \begin{equation}
  \cL_{al}=-ig_l\frac{m_l }{v_a} a'\bar e_l\gamma_5  e_l,
  \end{equation} 
  from which we define the Yukawa coupling 
 \begin{equation}
  h_{al} = g_l{m_l \over v_a}.
  \label{e:yukel}
  \end{equation}
It is useful for later calculations to
  express the couplings in terms of $\tan\beta$, where
  \begin{equation}
  \tan\beta\equiv{v_2\over v_1}.
  \end{equation}
  So for Model I:
  \begin{eqnarray}
  g_u &=& {v_2^2\over {v_1^2+v_2^2}}={\tan^2\beta\over{1+\tan^2\beta}}=\sin^2
  \beta, \\
  g_{d_j,e_l} &=& {v_1^2\over{v_1^2+v_2^2}}={1\over
  {1+\tan^2\beta}}=\cos^2\beta, \label{e:el1} \\
  g_{c,t} &=& -{v_1^2\over   
  {v_1^2+v_2^2}}=-{1\over{1+\tan^2\beta}}=-\cos^2\beta,
  \end{eqnarray}
  where $d_j=d,s,b$ and  $e_l=e,\mu,\tau$.  
  Using the same  notation we can work out the couplings for Model II,
  \begin{eqnarray}
  g_{u,c} &=& -{v_1^2\over  
  {v_1^2+v_2^2}}=-{1\over{1+\tan^2\beta}}=-\cos^2\beta, \\
  g_{d_j,e_l} &=& {v_1^2\over{v_1^2+v_2^2}}={1\over
  {1+\tan^2\beta}}=\cos^2\beta, \label{e:el2} \\  
  g_t &=& {v_2^2\over {v_1^2+v_2^2}}={\tan^2\beta\over{1+\tan^2\beta}}=\sin^2
  \beta,
  \end{eqnarray}
and for Model III:
  \begin{eqnarray}
  g_{u,t} &=& {v_2^2\over {v_1^2+v_2^2}}={\tan^2\beta\over{1+\tan^2\beta}}=
  \sin^2\beta,  \\ 
  g_{d_j,e_l} &=& {v_1^2\over{v_1^2+v_2^2}}={1\over
  {1+\tan^2\beta}}=\cos^2\beta, \label{e:el3} \\ 
  g_c &=& -{v_1^2\over {v_1^2+v_2^2}}=-{1\over{1+\tan^2\beta}}=-\cos^2
  \beta.
  \end{eqnarray}
  
 Lastly, we write down the interaction between the axion and photons, which 
 arises in the same way as for the axion-gluon interaction, through 
 the $a\gamma\gamma$ triangle diagram
  \begin{equation}
  \cL_{a\gamma}=N_e{e^2\over 32\pi^2 v_a}a'F_{\mu\nu}\tilde F^{\mu\nu},
  \end{equation}
where
  \begin{equation}
  N_e = \sum_j (Q_{Rj}-Q_{Lj}) e^2_j .
  \end{equation}
In this equation, $e_j$ is
  the electric  charge of the $j$-th fermion (with each colour quark counted 
  separately).
  
  It is important
  to note that below the QCD scale ($\sim  200$MeV), free quarks do not exist
  so one has to consider the effective  couplings of axions to nucleons, which
  arise from the axion mixing with $\pi^0$ and the 
  $\eta'$. It is exactly this mixing that allows us to calculate the axion mass
  and also has an effect on the coupling to two photons.
  
Standard current algebra techniques \cite{20,21} 
tell us that the physical axion $a$ has mass (including the strange quark)
\begin{equation}
m_a = \frac{m_\pi f_\pi}{v_a} N \left[\frac{z}{(1+z)(1+z+w)}\right]^{1/2}, 
\end{equation}
and that its coupling to photons is
  \begin{equation}
  \cL_{a\gamma\gamma}={1\over 4}h_{a\gamma} a F\tilde F,
  \end{equation}
where
  \begin{equation}
  h_{a\gamma}={e^2\over{4\pi^2}}\left[{1\over{v_a}}N\left({N_e\over
  N}-\frac{2(4+z+w)}{3(1+z+w)}\right)\right],
  \end{equation}
and we define $z$ and $w$ as \cite{22}
$$ z \equiv \frac{m_u}{m_d}=0.553\pm 0.043,$$
$$ w \equiv \frac{m_u}{m_s}=0.029\pm 0.0043.$$
For  models I \& II one has $(N , N_e)=(1/2 , 4/3)$ and for  model  III $(N ,
  N_e)=(1 , 8/3)$ .  As we can see the ratio  ${N_e/N}=8/3$  for all of them.
  This ratio  holds both for the DFSZ and the KSVZ axion, as well as for 
  GUT axion models based on SU(5) \cite{21}.
  
The same current algebra techniques enable us to compute the axion-nucleon
coupling \cite{21}. First we write down the anomaly-free axion current, 
including the strange quark
 \begin{eqnarray}
 j^a_{\mu}&=& v_a\partial_{\mu}a+{1\over 2}\left(g_u-\frac{2}{1+z+w}N\right)
 \bar 
u\gamma_{\mu}\gamma_5 u+{1\over 2}\left(g_d-\frac{2z}{1+z+w}N\right)\bar d
\gamma_{\mu}\gamma_5 d+ \nonumber \\
 & & +{1\over 2}\left(g_s-\frac{2w}{1+z+w}N\right)\bar s\gamma_{\mu}\gamma_5 s.
\label{e:AxCur}
\end{eqnarray}
 To find the couplings with the nucleons, we take the axionic current
 in the effective theory of nucleons,
 \begin{equation}
 j^a_{\mu}={1\over 2}\bar\psi(g_0+g_3\tau_3)\gamma_{\mu}\gamma_5\psi,
 \end{equation}
 where $\tau_3$ is the usual Pauli matrix, and $\psi$ is the nucleon doublet
  \begin{displaymath}
  \psi=\left( \begin{array}{c}
                 p \\ n
                \end{array}
          \right)                   , 
  \end{displaymath}
  sandwich it between nucleon states, and compare with the expression 
  obtained by using Eq.\ (\ref{e:AxCur}).
We find that the isoscalar and isovector couplings are given by
  \begin{eqnarray} 
  g_0 &=& {1\over 6}(\Delta u+\Delta d-2\Delta s)\left(g_u+g_d-2g_s-
  \frac{1+z-2w}{1+z+w}2N\right)+ \nonumber \\
     & &+{1\over 3}(\Delta u+\Delta d+\Delta s)(g_u+g_d+g_s-2N), \\
  g_3&=&{1\over 2}(\Delta u-\Delta d)\left(g_u-g_d-\frac{1-z}{1+z+w}2N\right),
  \end{eqnarray}
  where $s_{\mu}\Delta q\equiv\langle\bar\psi|\bar q\gamma_{\mu}\gamma_5q|\psi
  \rangle$, with $s_{\mu}$ being the spin of the nucleon. 
  We can find values 
for the $\Delta q$'s by combining the measurement of the proton 
spin structure function by the E143 collaboration \cite{23},
which gives $\Delta u + \Delta d + \Delta s = 0.27\pm 0.10$, with 
quark model and SU(3) flavour symmetry predictions for 
nucleon and hyperon couplings. We find 
\begin{eqnarray}
\Delta u &=& 0.813\pm 0.072, \nonumber \\
\Delta d &=& -0.444\pm 0.072, \nonumber \\
\Delta s &=& -0.10\pm 0.04. \nonumber
\end{eqnarray}
  Translating these 
  results into the language of the effective lagrangian approach we find
  \begin{equation}
  \cL_{a\bar\psi\psi}=-i{m_{\psi}\over v_a}a\bar\psi(g_0+g_3\tau_3)\gamma_5\psi.
  \end{equation}

  It is now straightforward to find the Yukawa couplings for
  the axion-proton and axion-neutron interactions.  For Model I,
  \begin{eqnarray}
  h_{ap}&=&\frac{m_p}{v_a}[-(0.678\pm 0.038)\cos 2\beta-(0.226\pm 0.058)], \\ 
  h_{an}&=&\frac{m_n}{v_a}[(0.578\pm 0.038)\cos 2\beta+(0.136\pm 0.058)],
  \end{eqnarray}
  for Model II,
  \begin{eqnarray}
  h_{ap}&=&\frac{m_p}{v_a}[-(0.678\pm 0.038)\cos 2\beta-(1.039\pm 0.058)], \\ 
  h_{an}&=&\frac{m_n}{v_a}[(0.578\pm 0.038)\cos 2\beta+(0.579\pm 0.058)],
  \end{eqnarray}
  and Model III
  \begin{eqnarray}
  h_{ap}&=&\frac{m_p}{v_a}[-(0.678\pm 0.038)\cos 2\beta-(0.587\pm 0.096)], \\ 
  h_{an}&=&\frac{m_n}{v_a}[(0.578\pm 0.038)\cos 2\beta+(0.137\pm 0.096)]. 
  \end{eqnarray}
 Finally the Yukawa couplings for the DFSZ axion are
 \begin{eqnarray}
 h_{ap}&=&\frac{m_p}{v_a}[-(1.357\pm 0.077)\cos 2\beta-(1.896\pm 0.275)], \\
 h_{an}&=&\frac{m_n}{v_a}[(1.157\pm 0.077)\cos 2\beta+(0.276\pm 0.275)].
 \end{eqnarray}
  
\section{Astrophysical constraints}

    We are now  interested  in seeing how the  astrophysical  limits  serve to
  constrain  the  couplings  and hence the PQ scale  $v_a$.  In our case these
  constraints  come  exclusively from the application of energy loss arguments
  to stars  \cite{11}.  According to it, if there are low mass  particles like
  neutrinos or novel  particles  interacting  weakly with matter and radiation
  like axions, they can be produced in large numbers in stellar interiors
  and can  afterwards  escape  freely.  In this way the stars are  drained  of
  energy and so alter their standard evolutionary course.

    We can now proceed to the  following  step,  which is  checking  how these
  three models behave under  constraints  coming from  astrophysics.  It would
  also be useful to compare the results  with the DFSZ model, mainly for two  
  reasons.
  First of all, it has significant similarities with our models, especially in
  the way that  fermions  acquire  their PQ  charges.  Secondly,  it is a well
  studied  and  established  axion model and all  existing  constraints  refer
  mostly to it.

  We begin by estimating the bounds on the axion  couplings to electrons. 
  The most  restrictive  bound comes from the so called  {\it helium  ignition
  argument} in red giants \cite{11,24}, which states that if a red giant
  produces
  a large number of  neutrinos or other  weakly  interacting  particles,  then
  helium  will  ignite in a much later time  because  larger  density  will be
  required.  For sufficiently light particles the energy loss rate would be so
  large,  that  helium  would not ignite at all, so the stars  would  directly
  become white dwarfs.  The limit is \cite{24}
  \begin{equation}
  h_{ae}<2.5\times 10^{-13}.
  \label{e:elcon}
  \end{equation}
  So according to equations (\ref{e:yukel}), (\ref{e:el1}), (\ref{e:el2}),
  (\ref{e:el3}) and (\ref{e:elcon}) we get the following lower limit for the
  axion decay constant:
  \begin{equation}
  v_a>2\times 10^9\cos^2\beta\ \textrm{GeV}, \\
  \end{equation}
  which holds for all models.  Another  important  interaction  axions have is
  the one with photons.  For the models under discussion it was first examined
  by Cheng, Geng \& Ni  \cite{25}.  The
  limit set on the  constraint  for the  axion-photon  coupling  is  \cite{26}
  \begin{equation}
  h_{a\gamma}<1\times 10^{-11}.
  \end{equation}
  From this and equation (24) is straightforward to conclude that
  \begin{equation}
  \hspace{4cm} v_a>8.5\times10^7\ \textrm{GeV},\hspace{2cm}\textrm{(Models\ I, 
II)} \\
  \end{equation}
  and 
  \begin{equation}
  \hspace{3.6cm} v_a>1.7\times 10^8\ \textrm{GeV},\hspace{2cm}\textrm{(Model\ 
III)} \\
  \end{equation}
  The  strictest  bound though  arises from the axion  coupling to  nucleons.
  The most  restrictive  value for this  constraint  come  from the  supernova
  SN1987a neutrino signal  \cite{27}.  The general  argument  underlying these
  kinds of  measurements is as follows.  We assume that we have some new light
  particles,  produced  in  the  interior  of  a  neutron  star,  more  weakly
  interacting than neutrinos e.g axions, and their Yukawa coupling noted 
  as $h_{a\psi}$, where $\psi$ is the nucleon doublet.  There are two 
  possibilities.  If  $h_{a\psi}$ is too small,
  then axions  cannot be trapped in the interior of the star.  Their mean free
  path is bigger  than the radius of the star and so they can  escape  freely.
  In this case axions are produced  inside the whole volume of the star and so
  the axion flux would be $L_a\propto  h_{a\psi}^2$.  The reason for this is 
  that
  axion  production would be dominated by processes like axion  bremsstrahlung
  from nucleons i.e  $n+p\longrightarrow  n+p+a$.  Hence, $L_a$ increases with
  increasing  $h_{a\psi}$  and eventually the axion flux equals the 
  neutrino flux $L_\nu$. Let this value of the coupling be $h_{min}$.
  On the other hand, for larger $h_{a\psi}$,  then  axions
  can be trapped and thermalised, in which case they are emitted from 
  a sphere of radius  $R_a$.  The stronger the coupling, the larger $R_a$ is.
  With blackbody  surface  emission,  $L_a\propto
  R_a^2  T^4(R_a)$, where $T(R_a)$ is the temperature at radius $R_a$.  
  For a nascent  neutron  star,  $R^2  T^4(R)$ is a rapidly
  decreasing   function   of  $R$.  
  We require that $L_a<L_\nu$, and thus $R_a > R_\nu$: 
  thus there is a value   of
  $h_{a\psi}$, say $h_{max}$,  above which the axion flux drops below the
  neutrino flux again.  Hence the range  $h_{min}<h_{a\psi}<h_{max}$ is
  excluded because 
  the axion flux would  dominate the one coming from the neutrinos. If 
  one is to remain conservative, many
  body effects should be taken into account,  in which case 
  the limit from this constraint is 
  \cite{28}
  \begin{equation}
  (h_{ap}^2+2h_{an}^2)^{1/2}<2.85\times 10^{-10}.
  \end{equation}
  This constraint puts bounds on $v_a$, which we write as
  \begin{equation}
  v_a > (0.33\times 10^{10})(A\cos^22\beta+B\cos2\beta+C)^{1/2}\ \mathrm{GeV}.
  \label{e:vbound}
  \end{equation}
 The values of $A$, $B$ and $C$, and their uncertainties (which arise 
 from the uncertainties in $\Delta q$, $z$, and $w$) are displayed in
 Table 1.
  
 For completeness, we also give the bounds on the DFSZ model 
 originating from the electron and photon couplings: 
  \begin{eqnarray}
 v_a &>& 4\times10^9\cos^2\beta\ \textrm{GeV}, \\
 v_a &>& 5.1\times 10^8\ \textrm{GeV}, 
  \end{eqnarray}
It is
clear that the lower bound on the axion scale from the red 
giant constraint (which operates on the axion-electron coupling) 
is relaxed by a factor of about 2.  The axion-photon coupling 
depends on $v_a/N$, as does the mass of the axion and the 
cosmological constraint in inflationary scenarios (in the axion
string scenario, $N=1/2$).  These particular constraints on the ratio 
$v_a/N$,  are therefore 
the same for our variant axion models as for DFSZ.

On the other hand, 
  it is not so straightforward to compare the
  nucleon constraints. For one to see how these bounds affect $v_a$, 
we  plot graphs of $v_a$ against $\beta$ for every model including the DFSZ.
   One can  easily  check from (Fig. 1) 
   that the lower bound is shifted downwards
  by a factor of 1.4--35 
  compared to that of the DFSZ axion. We have also displayed how the 
  uncertainties in the nucloeon couplings are propagated into uncertainties 
  in the bounds.

  \section{Conclusions}

  Although  the axion  solution  to the  strong CP problem  is one of the most
  physically  appealing, axions  themselves face a great problem.  Despite the
  fact that they interact very weakly with matter and so are very difficult to
  track,   particle   physics   experiments    together   with   astrophysical
  considerations  and  cosmology  have  managed  to constrain axion models 
  significantly.  In this  paper we have found the constraints
 on  for axion models with non-standard couplings to 
  quarks and leptons, using data from E143 to determine the values of
  the nucleon couplings, which provide the strongest constraint. 
We find that the bounds are generally weakened, as the nucleon couplings 
in our variant axion models are smaller. The most spectacular effect is
for Model I near $\beta \sim \pi/4$: the bound dips to about 
$v_a > 2\times 10^8$ GeV, about a factor 35 less than the DFSZ value.
  Models I and II have 
  the  desirable  feature that the QCD anomaly coefficient $N={1\over 2}$, 
  which means that they have no domain wall problem, and therefore are
  viable models for an axion string scenario.  For Model I, the
  lower bound on $v_a$ can dip to $2\times10^8$ GeV for values of 
  $\beta$ near $\pi/4$. Recalling the upper bound on $v_a$ in the 
  axion string scenario, $v_a < 2.6\times10^{11}$ GeV for $H_0 = 50$ 
  km s$^{-1}$ Mpc$^{-1}$ \cite{14}, we see that the `window' for this
  axion string scenario is actually quite large.  

In an inflationary scenario, the cosmological upper bound is on 
$v_a/N$, and so the upper bound on $v_a$ itself is reduced by a factor 
of 6 for Models I and II, and a factor 3 for Model III.  

\acknowledgements

M.H.\ is supported by PPARC Advanced Fellowship  
B/93/AF/1642 and by PPARC grant  GR/K55967.

\begin{table}[ht]
\begin{tabular}{|l|ccc|}
Model & $A$ & $B$ & $C$ \\
\hline
I & $1.13\pm0.14$ & $0.62\pm0.18$ & $0.09\pm0.06$ \\
II& $1.13\pm0.14$ & $2.75\pm0.31$ & $1.75\pm 0.26$\\
III&$1.13\pm0.14$ & $1.11\pm0.31$ & $0.38\pm 0.17$\\
DFSZ&$4.52\pm0.14$&$6.42\pm1.76$ &$ 3.75\pm1.35$\\
\end{tabular}

\caption{Table of values of coefficients in Eq.\ (\ref{e:vbound}), 
 with experminetal uncertainties, for the three variant axion models 
 considered, compared with the values for the canonical DFSZ model.}
\end{table}

\begin{figure}[ht]
  \centering
  \scalebox {0.75} {\includegraphics {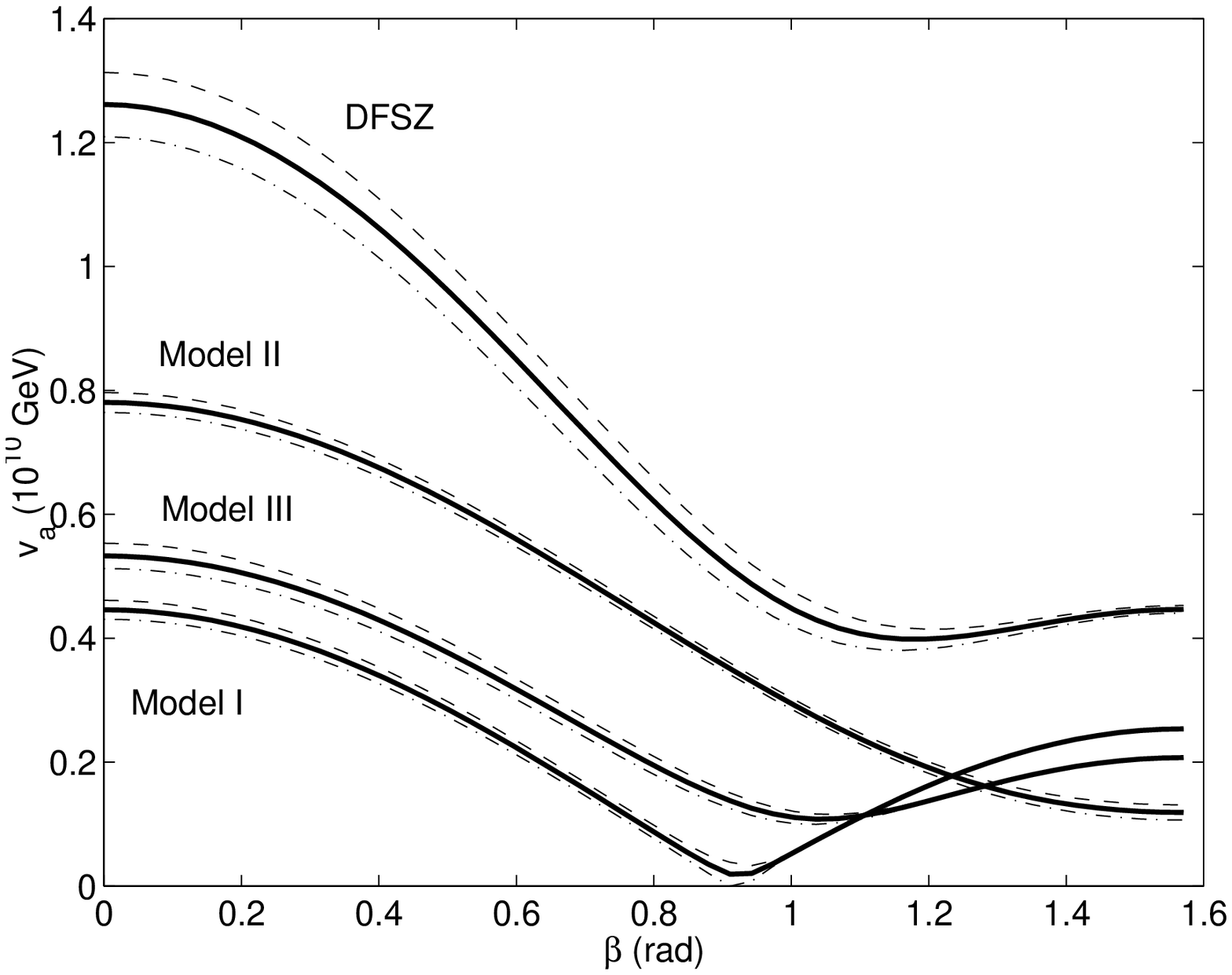}}
  
  \caption {\label{figure 1}The lower bound on the Peccei-Quinn 
  symmetry-breaking scale $v_a$ for variant axion models described
  in the text and the DFSZ model, plotted as a function of $\beta$. 
  Here, $\tan \beta = v_2/v_1$, where $v_1$ and $v_2$ are the vacuum 
  expectation values of the two higgs fields giving masses to the
  fermions. The dashed and dot-dashed lines indicate the uncertainties
  arising from the nucleon couplings. }
  \end{figure}


\begin{references} 
\bibitem[*]{mhaddress}Electronic address: 
m.b.hindmarsh@sussex.ac.uk
\bibitem[\dagger]{pmaddress}Electronic address: 
p.moulatsiotis@sussex.ac.uk
\bibitem{1}  A.A.  Belavin, A.M.  Polyakov,  A.S.  Schwartz and Y.S.
  Tyupkin, {\it Phys.  Lett.} {\bf B59}, 85 (1975).
\bibitem{2} G.  't Hooft, {\it Phys.  Rev.  Lett.} {\bf 37}, 8 (1976); G.  't
  Hooft, {\it Phys.  Rev.}  {\bf D14}, 3432 (1976).
\bibitem{3} S.L.  Glashow, {\it Hadrons and their Interactions}, Proc.
  1967 Int.  School of Physics "Ettore Majorana", Academic Press, New York; S.
  Weinberg, {\it Phys.  Rev.}  {\bf D11}, 3583 (1975).
\bibitem{4} C.G.  Callan, R.F.  Dashen and D.J.  Gross, {\it Phys.
  Lett.}  {\bf B63}, 334 (1976).
\bibitem{5}  R.  Jackiw  and C.  Rebbi, {\it Phys.  Rev.  Lett.}  {\bf  37},  
  177  (1976).
\bibitem{6} R.  Crewther, P.  Di Vecchia, G.  Veneziano and E.  Witten, {\it
  Phys.  Lett.}  {\bf B88}, 123 (1979); V.  Baluni, {\it Phys.  Rev.}  {\bf
  D19}, 2227 (1979); J.  Bijens, H.  Sonoda and M.B.  Wise, {\it Nucl.
  Phys.}  {\bf B261}, 185 (1985).
\bibitem{7} K.F. Smith et al., {\it Phys. Lett.} {\bf B234}, 191 (1990);
  I.S.  Altarev et al., {\it Phys.  Lett.}  {\bf B276}, 242 (1992).
\bibitem{8} R.D.  Peccei and H.R.  Quinn, {\it Phys.  Rev.  Lett.}  {\bf
  38}, 1440 (1977); R.D.  Peccei and H.R.  Quinn, {\it Phys.  Rev.}  {\bf
  D16}, 1791 (1977).
\bibitem{9} S. Weinberg, {\it Phys. Rev. Lett.} {\bf 40}, 223 (1978);
  F. Wilczek, {\it Phys. Rev. Lett.} {\bf 40}, 279 (1978).
\bibitem{10} M.  Sivertz et al., {\it Phys.  Rev.}  {\bf D26}, 717
  (1982); G.  Edwards et al., {\it Phys.  Rev.  Lett.}  {\bf 48}, 903
  (1982); M.S.  Salam et al., {\it Phys.  Rev.}  {\bf D27}, 1665
  (1983); B.  Niczyporuk et al., {\it Z.  Phys.}  {\bf C17}, 197 (1983);
  Y.  Asano et al., {\it Phys.  Lett.}  {\bf B107}, 159 (1982); Y.
  Asano et al., {\it Phys.  Lett.}  {\bf B113}, 195 (1982); G.  Mageras
  et al., {\it Phys.  Rev.  Lett.}  {\bf 56}, 2672 (1986); T.  Bowcock
  et al., {\it Phys. Rev.  Lett.}  {\bf 56}, 2676 (1986).
\bibitem{11} G.G.  Raffelt, {\it Phys.  Rep.}  {\bf 198}, 1 (1990) and
  references therein.
\bibitem{12} J.E.  Kim, {\it Phys.  Rev.  Lett.}  {\bf 43}, 103 (1979); 
  M.A.  Shifman, V.I.  Vainstein and V.I.  Zakharov, {\it Nucl.  Phys.}
  {\bf B166}, 4933 (1980).
\bibitem{13} J.  Preskill, M.B.  Wise and F.  Wilczek, {\it Phys.  Lett.}
  {\bf B120}, 127 (1983); L.F.  Abbott and P.  Sikivie, {\it Phys.  Lett.}
  {\bf B120}, 133 (1983); M.  Dine and W.  Fischler, {\it Phys.  Lett.}  {\bf
  B120}, 137 (1983); M.S.  Turner, {\it Phys.  Rev.}  {\bf D33}, 889
  (1986).
\bibitem{14} R.A.  Battye and E.P.S.  Shellard, {\it Phys.  Rev.
  Lett.}  {\bf 73}, 2954 (1994), {\it ibid} {\bf 76}, 2203 (1996); 
  R.A.  Battye and E.P.S.  Shellard, {\it Critique of the Sources of Dark
  Matter in the Universe}, Proc.  1994 Int.  Symposium at UCLA.
\bibitem{15} P. Sikivie, {\it Phys. Rev. Lett.} {\bf 48}, 1156 (1982).
\bibitem{16} R.D.  Peccei, T.T.  Wu and T.  Yanagida, {\it Phys.  Lett.}
  {\bf B172}, 435 (1986).
\bibitem{17} L.M.  Krauss and F.  Wilczek, {\it Phys.  Lett.}  {\bf B173},
  189 (1986).
\bibitem{18} A.P.  Zhitnitskii, {\it Sov.  J.  Nucl.  Phys.}  {\bf 31},
  260 (1980); M.  Dine, W.  Fischler and M.  Srednicki, {\it Phys.  Lett.}
  {\bf B104}, 199 (1981).
\bibitem{19} C.Q.  Geng and J.N.  Ng, {\it Phys.  Rev.}  {\bf D39}, 1449
  (1989).
\bibitem{20} W.A. Bardeen and S.-H.H. Tye, {\it Phys. Lett.} {\bf B74},
  229 (1978); J. Kandaswamy, P. Salomonson and J. Schechter, {\it Phys. Rev.}
  {\bf D17}, 3051 (1978).
\bibitem{21} D.B.  Kaplan, {\it Nucl.  Phys.}  {\bf B260}, 215 (1985); M.
  Srednicki, {\it Nucl.  Phys.}  {\bf B260}, 689 (1985).
\bibitem{22} H. Leutwyler, {\it Phys. Lett.} {\bf B378}, 313 (1996).
\bibitem{23} K. Abe et al., {\it Phys. Rev. Lett.} {\bf 74}, 346 (1995).
\bibitem{24} G.G.  Raffelt and A.  Weiss, {\it Phys.  Rev.}  {\bf D51},
  1495 (1995).
\bibitem{25} S.L.  Cheng, C.Q.  Geng and W.-T.  Ni, {\it Phys.Rev.}
  {\bf D52}, 3132 (1995).
\bibitem{26} J.W.  Brockway, E.D.  Carlson and G.G.  Raffelt, {\it
  Phys.  Lett.}  {\bf B383}, 439 (1996).
\bibitem{27} GG.  Raffelt and D.  Seckel, {\it Phys.  Rev.  Lett.}  {\bf
  60}, 1793 (1988); M.S.  Turner, {\it Phys.  Rev.  Lett.}  {\bf 60}, 1797
  (1988); R.  Mayle et al., {\it Phys.  Lett.}  {\bf B203}, 188 (1988);
  R.  Mayle et al., {\it Phys.  Lett.}  {\bf B219}, 515 (1989); A.
  Burrows, M.S.  Turner and R.P.  Brinkmann, {\it Phys.  Rev.}  {\bf D39},
  1020 (1989).
\bibitem{28} W. Keil et al., astro-ph/9612222. To be published in {\it Phys.
  Rev. D}.
\end{references}
  \end{document}